\documentclass[12pt]{article}

\usepackage{graphicx}
\usepackage{lscape}
\begin{document}
\baselineskip=24pt

\title{A model of cardiac tissue as a conductive system with
interacting pacemakers and refractory time}

\author{Alexander Loskutov, Sergei Rybalko and Ekaterina Zhuchkova}
\date{}
\maketitle

\begin{center}
{\small Physics Faculty, Moscow State University,
Leninskie gory, 119992 Moscow, Russia}

E-mail: loskutov@moldyn.phys.msu.ru
\end{center}

\begin{abstract}
A model of the heart tissue as a conductive system with two
interacting pacemakers and a refractory time, is proposed. In the
parametric space of the model the phase locking areas are
investigated in detail. Obtained results allow us to predict the
behaviour of excitable systems with two pacemakers depending on
the type and intensity of their interaction and the initial phase.
Comparison of the described phenomena with intrinsic pathologies
of cardiac rhythms is presented.
\end{abstract}

\vspace{1cm}
{\bf Short title:} A model of cardiac tissue  with
interacting pacemakers

\section{Introduction}

One of the remarkable examples of excitable media is the cardiac
tissue. Because the stability of its behaviour is essential for
living creatures, investigations of processes occurring in the
cardiac muscle attract a considerable interest of various
scientists. Owing to a great complexity and a wide variety of the
processes taking place in the heart, it can be considered with
several points of view. One of them relies upon the interpretation
of the cardiac tissue as an active conductive system. Then, the
cardiac rhythms are described on the basis of the dynamical
systems theory (see, for example [Glass {\it et al.}, 2002]).

The  excitation wave in the cardiac tissue originates in the
sinoatrial node (SA) and spreads rapidly in succession over the
right atrium, the left atrium, then the atrioventricular node
(AV), bundle of His and Purkinje fibers, and finally to the walls
of the right and left ventricles. The normal rhythm of the heart
is determined by the activity of the SA node which is called the
leading pacemaker (a source of concentric excitation waves) or the
first order driver of the rhythm. In addition to the SA node
cells, the other parts of the cardiac conductive system reveal an
automaticity. So, the second order driver of the rhythm is located
in the AV conjunction. The Purkinje fibers are the rhythm driver
of the third order. Moreover, for some pathological states of the
heart arising of ectopic pacemakers is typical.

Appearance of several excitation sources leads to various
disorders in the cardiac rhythm, i.e. arrhythmias [Schamorth,
1980; Marriot \& Conover, 1983; Zipes \&  Jalife, 1985; Winfree, 1987; Glass \&
Mackey, 1988]. Because arrhythmias are dangerous diseases of the
heart, their investigations have a great importance. Analysis of
the complex cardiac rhythms on the basis of their interpretation
as chaotic phenomena can give a clue to the problem of
controllability of the complex cardiac dynamics and removing the
cardiac tissue to the required regime [Goldberger \& Rigney, 1988;
Goldberger, 1990; Garfinkel {\it et al.}, 1992].

As known, some arrhythmias can be presented by the interaction of
spontaneous nonlinear sources [Glass {\it et al.}, 1983;
Glass \& Mackey, 1988; Kremmydas {\it et al.}, 1996]. In
some cases, a model of impulse systems describing certain types of
such arrhythmias is constructed in the framework of the theory of
dynamical systems. This model is represented by coupled
low-dimensional maps (circle maps) which can exhibit a complex
dynamics. Such an approach is based on the fact that experiments
on periodically stimulated cardiac cells are in a close agreement
with the dynamics predicted by one-dimensional circle maps
[Courtemanche {\it et al.}, 1989].

In our investigations, a quite general model of two nonlinear
coupled oscillators describing certain types of cardiac
arrhythmias is constructed. The model turns out to be a universal
one in the sense that it does not depend on the specific form of
interactions, i.e. on the phase response curve (PRC). The
experimentally obtained PRC is approximated by a certain
polynomial function with plateau. This plateau describes a
refractory time when the system does not respond to an external
action. Note that the refractory time plays an important role for
the normal cardiac functioning. For example, the refractoriness
extends over almost the whole period of the cardiac contraction
protecting the myocardium from premature heartbeats caused by an
external perturbation. The refractoriness provides also the normal
sequence of an excitation propagation in the heart tissue and the
electrical stability of the myocardium [Marriot \& Conover, 1983;
Zipes \&  Jalife, 1985; Winfree, 1987; Glass \& Mackey, 1988]. In the proposed
model, taking into consideration the refractory time possible
areas of the phase lockings are investigated. Phenomena of the
splitting of resonance tongues and superposition of the
synchronization areas are found. Using the obtained results we can
define dynamics of the excitable media with two active pacemakers
depending on the type and intensity of their interaction and the
initial phase difference. Moreover, generalizing the principles of
our construction one can develop a quite general theory of
excitable media with interacting pacemakers under external
actions. This fact has a great practical importance because admits
to realize the control of the cardiac rhythms by external stimuli.

\section{Heart Tissue as a Dynamical System}

In certain cases cardiac arrhythmias can be described as an
interaction of two spontaneously oscillating nonlinear sources.
This interaction  can be considered as an influence of some
external periodic perturbation on a nonlinear oscillator (with the
constant amplitude and frequency). So, for the description of such
a situation it is possible to use the well-known circle map
[Glass {\it et al.}, 1983; Bub \& Glass, 1994;
Kremmydas {\it et al.}, 1996]:
\[
x_{n + 1} = x_{n} + f\left( {x_{n}}  \right)
\quad\left( \textrm{mod \ 1} \right),
\]
where $x_{n}$ is a phase difference in oscillators and the function
$f\left({x}\right)$ determines a change in the phase after action of
stimulus. This function is called a phase response curve (PRC).

One of the most important characteristics of the circle map is a
rotation number $\rho$. It is defined as follows:
$$
\rho=\lim_{n\rightarrow\infty}\frac{x_n-x_0}{n}\ .
$$
For stable phase locking $N:M$ the rotation number is rational,
$\rho=N/M$. If it is irrational the system behaviour is
quasiperiodic or chaotic.

Analysing dynamics of the constructed model based on
the circle map, it is necessary to find a proper analytical
approximation of the experimentally obtained phase response curve.
This allows us to investigate the basic features of the behaviour
of the considered system.

Experiments on the recording of phase shifts have  been carried
out for a quite large number of various systems. First of all, we
are interested in the PRC experimentally obtained from the
research of some cardiac tissues. In [Weidmann, 1961; Jalife \&
Moe, 1976] measurements of the cycle durations of the spontaneous
beating Purkinje fibres after stimulation by short electric
current pulses have been performed. The found  phase response
curve is shown in Fig.\ref{fig1} (dotted lines). Taking into
account this experimental material, it is possible to make the
following general conclusions [Weidmann, 1961; Jalife \& Moe,
1976; Glass {\it et al.}, 1986; Glass \& Mackey, 1988]:

\begin{itemize}
\item
after perturbation the rhythm is usually restored
(after some transient time)
with the same frequency and amplitude, but the phase is shifted;
\item
depending on a phase the single input can lead to either increasing or
decreasing of the period of a perturbed cycle;
\item
at some amplitudes of stimulus the obvious breaks appear.
\end{itemize}

The basic feature of any approximation of the PRC is the
dependence on two physical parameters: on the amplitude of
stimulus and the input phase. In the ideal case the other
(so-called "internal") parameters can be reduced to them.

Taking into account the polynomial function for the approximation
of the PRC, we construct a model of two \textit{mutually}
interacting impulse active oscillators.

\section{Analytical Model with a Mutual Influence of Impulses}

Let us consider the system of two  nonlinear interacting oscillators
(Fig.\ref{fig2}). Suppose that the pulse of the first oscillator with
period $T_{1} $ beats at $t_{n}$, and the pulse of the second
oscillator (with period $T_{2}$) beats at $\tau_{n}$. Then the
moments of time of the next appearance of the impulses are defined
as follows:
\[
\left\{ {{\begin{array}{*{20}c}
{t_{n + 1} = t_{n} + T_{1} ,} \hfill \\
\\
{\tau _{n + 1} = \tau _{n} + T_{2} .} \hfill \\
\end{array}} } \right.
\]
Now, taking into account the change in the period of the first
oscillator under the influence of the second impulse by the value
of $ \Delta_{1} \Bigl ({\left ({\tau _ {n} - t _ {n}}
\right)/T_{1}} \Bigr)$, one can get that $t _ {n +1} = t _ {n}+ T
_ {1}+ \Delta _ {1} \Bigl({\left ({\tau _ {n} - t _ {n}} \right)/T
_ {1}} \Bigr)$. By the same manner, for the second oscillator $
\tau _ {n+ 1} = \tau _ {n}+ T _ {2}+ \Delta _ {2} \Bigl({\left ({t
_ {n+ 1} - \tau _ {n}} \right)/T _ {2}} \Bigr) $. Dividing these
expressions by $T _ {1} $ we arrive at the corresponding values
for the phases:
\[
\left\{ {{\begin{array}{*{20}c}
{\varphi _{n + 1} = \varphi _{n} +
\displaystyle\frac{{1}}{{T_{1}} }\Delta _{1} \left(
{\delta _{n} - \varphi _{n}}  \right),
\quad \quad \quad \quad \quad \quad
\quad \quad \quad \quad}  \hfill \\
\\
{\delta _{n + 1} = \delta _{n} + \displaystyle
\frac{{T_{2}} }{{T_{1}} } +
\displaystyle\frac{{1}}{{T_{1}} }\Delta _{2}
\left( {\displaystyle\frac{{t_{n}} }{{T_{2}} } +
\displaystyle\frac{{T_{1}} }{{T_{2}} } +
\displaystyle\frac{{1}}{{T_{2}} }
\Delta _{1} \left( {\delta_{n} - \varphi _{n}}\right)-
\displaystyle\frac{{\tau _{n}} }{{T_{2}} }} \right).}
\hfill \\
\end{array}} } \right.
\]
Here $ \varphi _ {n}=t _ {n}/T _ {1} $ is a phase of the first
perturbed oscillator with respect to the unperturbed one (with
period $T _ {1} $), and $\delta _ {n} = \tau _ {n}/T _ {1} $ is
the phase of the second perturbed oscillator with respect to the
same first oscillator with period $T _ {1} $. Using parameters $a
= T _ {2}/T _ {1} $ and $ \Delta _ {1}/T _ {1} = f _ {1} $, $
\Delta_{2}/T _ {1} = f _ {2}$ one can write:
\[
\left\{ {{\begin{array}{*{20}c}
{\varphi _{n + 1} = \varphi _{n} + f_{1} \left( {\delta _{n} -
\varphi _{n}} \right),\quad \quad \quad \quad \quad \quad
\quad \quad \quad}  \hfill \\
\\
{\delta _{n + 1} = \delta _{n} + a + f_{2} \left(
{\displaystyle\frac{{1}}{{a}}\left({\varphi _{n} +
1 + f_{1} \left( {\delta _{n} - \varphi _{n}}  \right) -
\delta _{n}}  \right)} \right).} \hfill \\
\end{array}} } \right.
\]
Now, omitting intermediate calculations, for the final expression
of the phase difference in the oscillators we get
\begin{equation}
\label{eq2}
x_{n + 1} = x_{n} + a + f_{2} \left( {\displaystyle
\frac{{1}}{{a}}\left( {1 + f_{1}\left( {x_{n}}\right)-
x_{n}}  \right)} \right) - f_{1} \left( {x_{n}}
\right)\quad \left( \textrm{mod \ 1} \right),
\end{equation}
where $x _ {n} = \delta _ {n} - \varphi _ {n} $.

It is obvious, that  the PRC changes the form depending on the
amplitude of the external stimulus. In the simplest case this
dependence can be considered as the multiplicative relation. Then
the phase response curves can be written as
\[
f_{1} = \gamma h\left( {x} \right),\quad \quad \quad f_{2} =
\varepsilon h\left( {x} \right),
\]
where $h\left ({x} \right) $ is a periodic function and
$h\left ({x+ 1} \right)= h\left ({x} \right) $.
In such an assumption, the expression
(\ref{eq2}) takes the form:
\begin{equation}
\label{eq3} x_{n + 1} = x_{n} + a + \varepsilon h \left(
{\displaystyle\frac{{1}}{{a}}\left( {1 + \gamma h\left( {x_{n}}
\right) - x_{n}}  \right)} \right) - \gamma h\left( {x_{n}}
\right)\quad \quad \left( \textrm{mod \ 1} \right).
\end{equation}

In the present paper we dwell on the investigations of the map
(\ref{eq3}) with a polynomial function $h\left({x}\right)$. The
obtained results are the continuation of our previous works
concerning modeling certain cardiac arrhythmias [Loskutov, 1994;
Loskutov {\it et al.}, 2002].

\section{Phase Diagrams for Unidirectional Coupling of Oscillators}

\begin{figure}[h]
  \centering
  \includegraphics[scale=0.8]{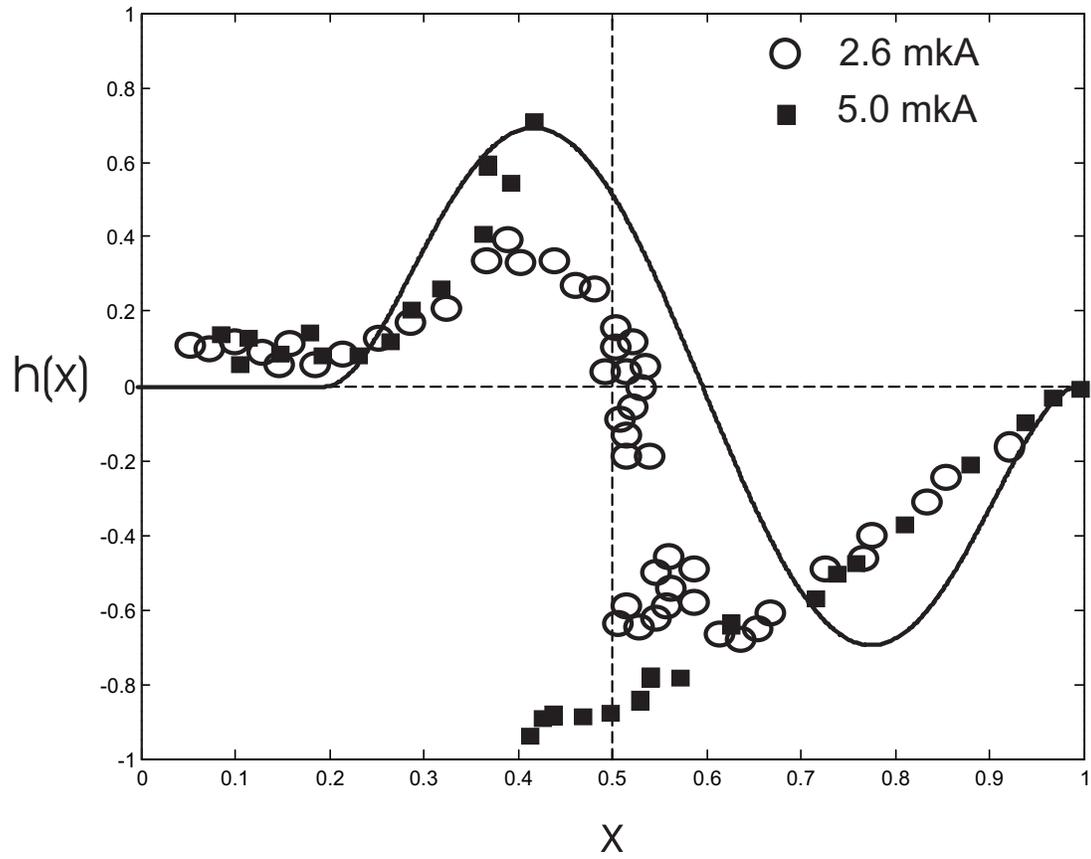}
   \caption{A. Loskutov, S. Rybalko \& E. Zhuchkova}
\label{fig1}
\end{figure}

\begin{figure}[h]
  \centering
  \includegraphics[scale=0.75]{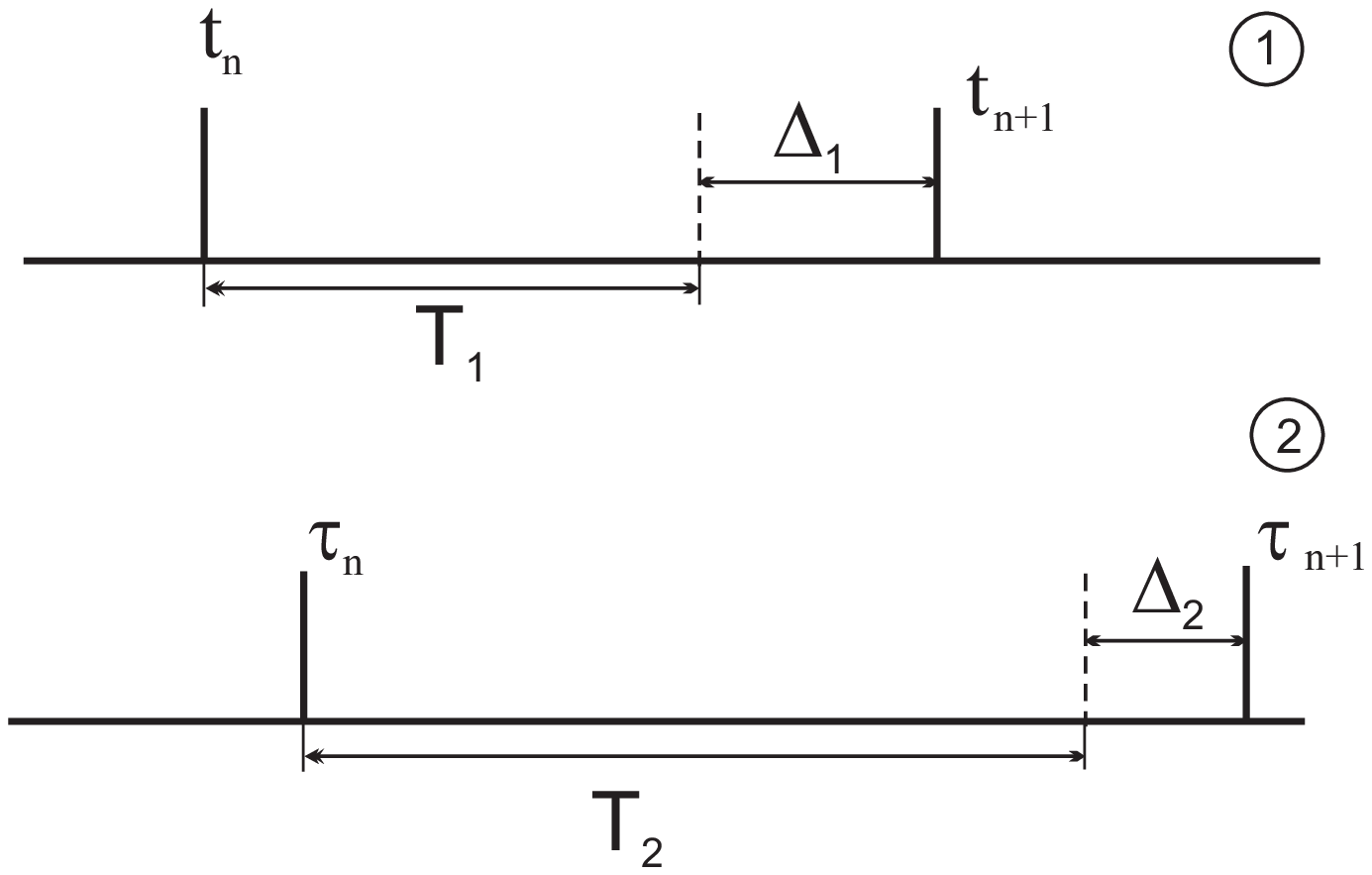}
   \caption{A. Loskutov, S. Rybalko \& E. Zhuchkova}
\label{fig2}
\end{figure}

\begin{figure}[h!]
  \centering
 \includegraphics[scale=1.2]{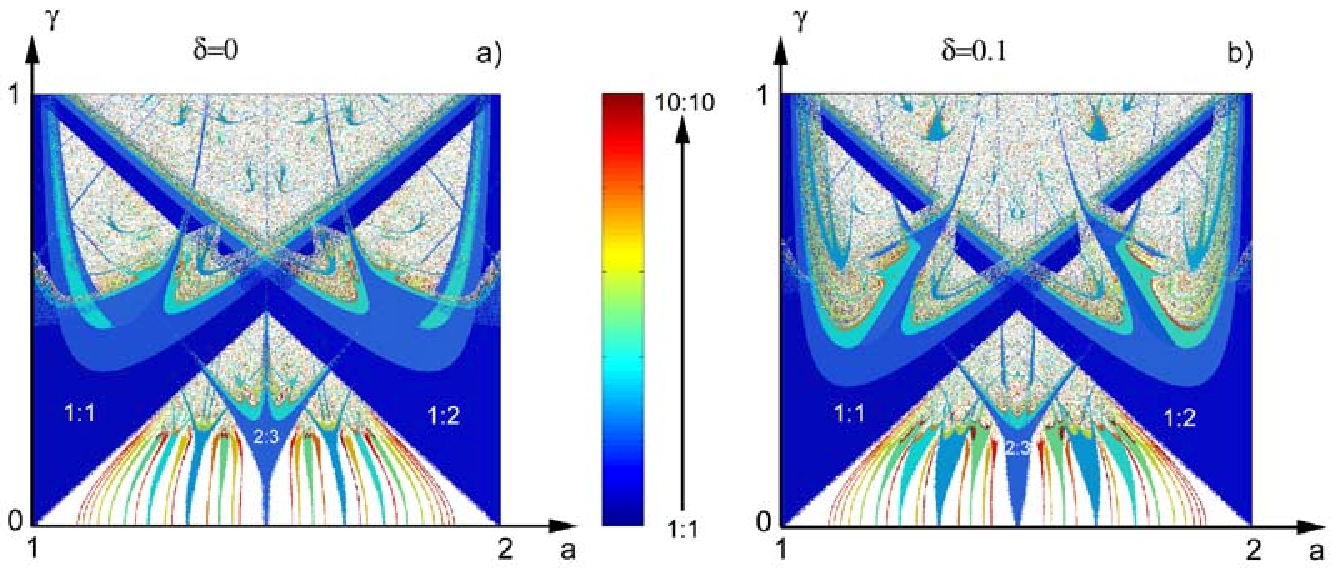}
 \caption{A. Loskutov, S. Rybalko \& E. Zhuchkova}
\label{fig4}
\end{figure}

First of all let us analyze the situation when the permanent
inputs act on the nonlinear oscillator, i.e. $f_{2}\left ({x}
\right) \equiv 0$ or $\varepsilon = 0$. As an analytical
approximation of the experimental curve in Fig.\ref{fig1}, let us
consider the following polynomial function:
\begin{equation}
h\left ({x}\right) =
C x ^ {2} \left ({\displaystyle\frac {{1}} {{2}} -
x} \right) \left ({1 - x} \right)^{2}.
\label{h}
\end{equation}
The normalizing factor $C$ we choose in such a way that the
amplitude of $h(x)$ is equal to $1$, so that $C=20\sqrt {5}$ (see
Fig.\ref{fig1}, solid line). Then taking into account the
refractory time $\delta$ the map (\ref{eq3}) we can write as
follows:
\begin{equation}
\label{eq5} x_{n + 1} = \left\{ {{\begin{array}{*{20}l} {x_{n} +
a,} & {0 \le x_{n} \le \delta},  & {\left( \textrm{mod \ 1}
\right),} \\ {x_{n} + a + C\gamma
h\left({\displaystyle\frac{x_n-\delta}{1-\delta}}\right),} &
{\delta < x_N \le 1}, & {\left(\textrm{mod \ 1} \right),} \\
\end{array}} } \right.
\end{equation}
where $h(\cdot)$ is determined by (\ref{h}). Now let us compare
several cases with different values of the refractory time. First
of all consider the case without the refractory period, i.e.
$\delta=0$. The phase locking regions in the parametric space
$\left({a,\gamma}\right)$ obtained by numerical analysis are shown
in Fig.\ref{fig4}a. Without loss of generality, in
this Figure we choose $a\in [1,2]$. Different colours define the phase
locking areas with the multiplicity $N:M$, where $N$ cycles of
external stimulus correspond to $M$ cycles of nonlinear oscillator.
One can see that "tales" of the main locking regions are slightly
splitted and overlap each other at large $\gamma$. Note that as
follows from the analysis of the system (\ref{eq5}) with
$\delta=0.1$ (Fig.\ref{fig4}b), introduction of the refractory
time leads to the extension of the phase locking areas and
significant splitting and overlapping their "tales".

\begin{figure}[h!]
  \centering
 \includegraphics[scale=1.2]{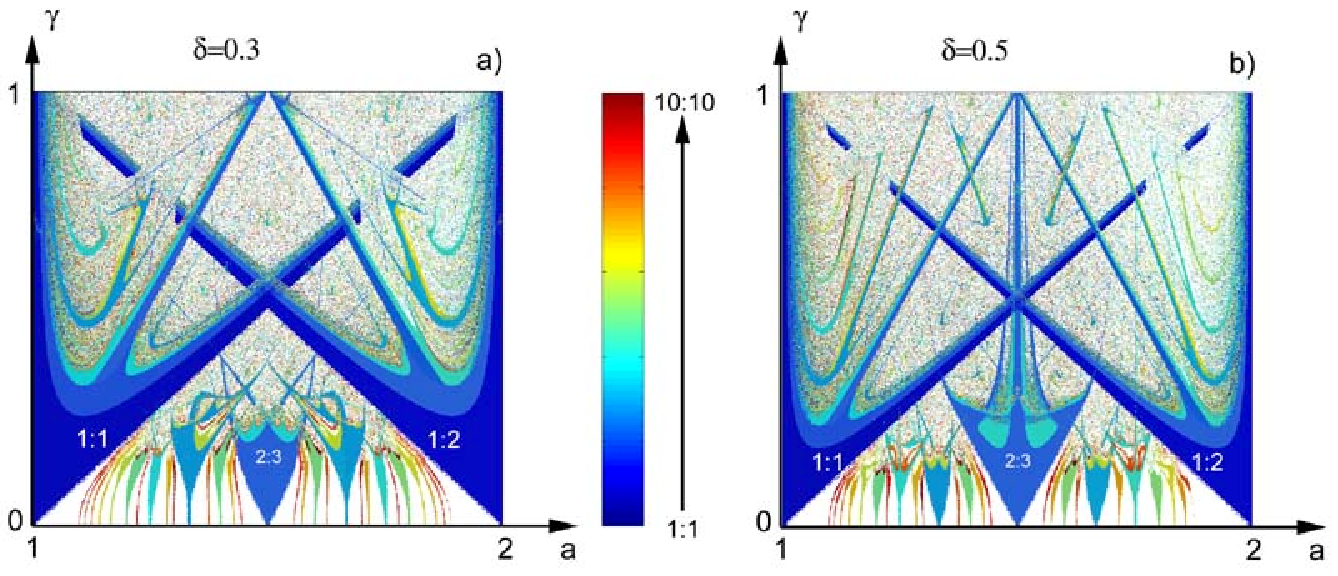}
  \caption{A. Loskutov, S. Rybalko \& E. Zhuchkova}
\label{fig5}
\end{figure}

In Fig.\ref{fig5}a the numerically constructed phase diagram in
the case of $\delta=0.3$ is presented. For the comparison, in the
given Figure the same $N:M$ stable phase lockings as in Fig.4 are
shown. One can see that when the value of the refractory time is
growing, the $2:3$ phase locking area is increasing with
simultaneous decreasing of the $1:1$ and $1:2$ areas.

\begin{figure}[h!]
  \centering
 \includegraphics[scale=1.2]{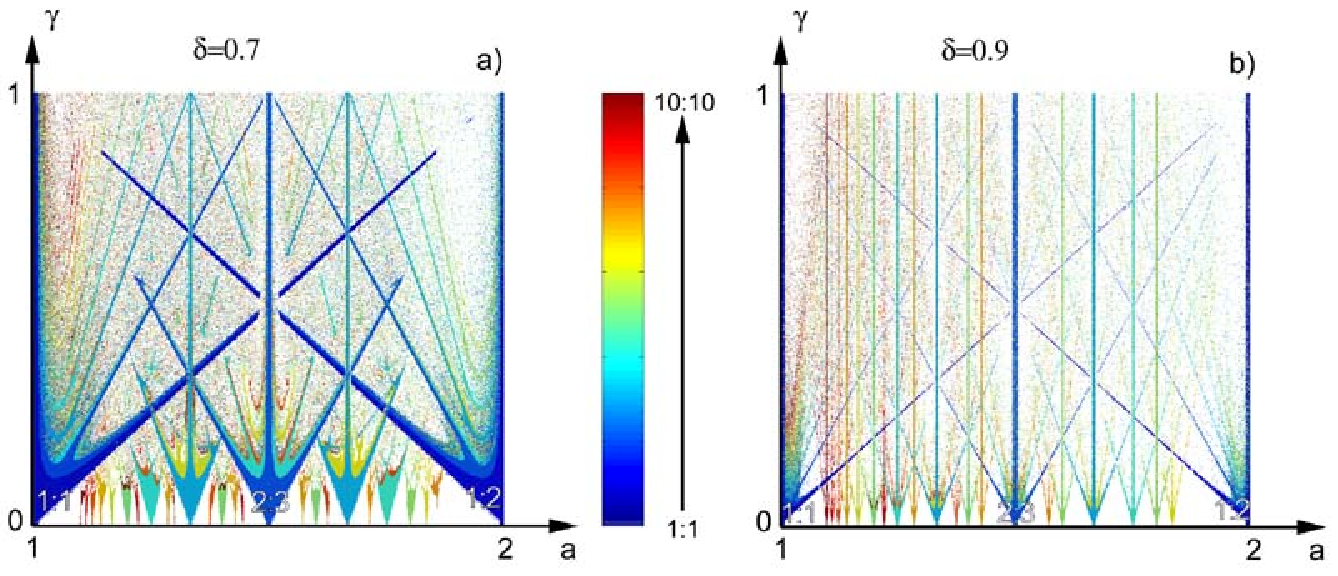}
  \caption{A. Loskutov, S. Rybalko \& E. Zhuchkova}
\label{fig6}
\end{figure}

The phase locking regions in the case of $\delta=0.5$ are shown in
Fig.\ref{fig5}b. This phase diagram is {\it qualitatively}
different from pictures considered above. The form of $2:3$ phase
locking area is stretched and looks like an arrow. The forms of
$3:4$ and $3:5$ regions also resemble arrows in the case of
$\delta=0.7$ (Fig.\ref{fig6}a). At $\delta=0.9$ all phase lockings
are degenerated into vertical lines. This situation is presented
in Fig.\ref{fig6}b. Note that in the case of $\delta=1$ (i.e. the
system does not respond to the external action) there is no any
dependence on the stimulus amplitude $\gamma$.

\section{Phase Diagrams for Systems with Mutual Interaction}

In this section the system (\ref{eq3}) at $\delta=0.1$ is
considered. The analysis is performed in the ($\gamma$,
$a$) and ($\gamma$, $\varepsilon$)--parametric spaces.

\subsection{Phase locking areas in the ($\gamma$,
$a$)--space}

\begin{figure}[h!]
  \centering
 \includegraphics[scale=1.2]{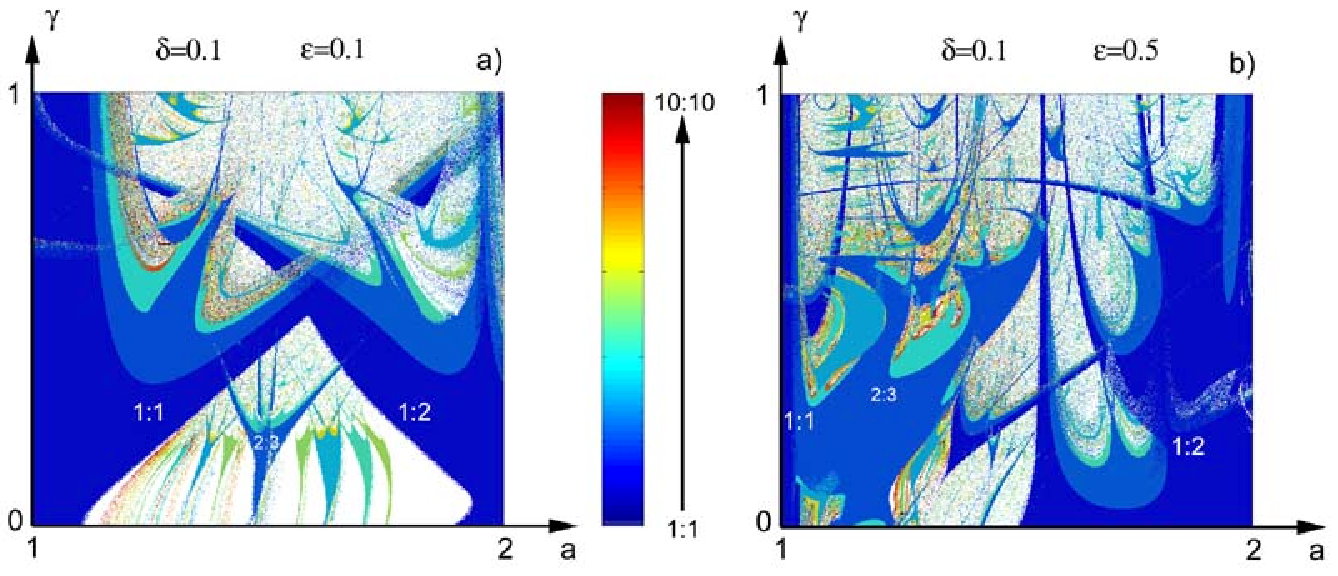}
 \caption{A. Loskutov, S. Rybalko \& E. Zhuchkova}
\label{fig7}
\end{figure}

Let us consider the case of a mutual interaction of two impulse
systems. Assume that the influence of the first oscillator on the
second one is small enough, for example, $\varepsilon = 0.1$. The
corresponding phase diagram displaying the possible behaviour
regimes of the system for $\delta=0.1$ is shown in
Fig.\ref{fig7}a. One can see that the mutual interaction leads to
the deformation and the splitting of the phase locking areas. Note
that even for small values of the amplitude of the second stimulus
$\gamma$, the overlapping of the main phase lockings takes place.
Thus, the system dynamics becomes multistable. This corresponds to
the situation when the limit state of the map depends on an
initial phase difference $x_{n}$. The growth of the refractory
time in the model with $\varepsilon = 0.1$ leads to more deep
distortion in the forms of the main tongues and disappearance of
the splitting areas. If, however, we increase the influence of the
first oscillator up to, for instance $\varepsilon = 0.5$, then one
can see a very complicated structure with {\it much more deep} deformation
of the main phase locking areas (see Fig.\ref{fig7}b). For
example, the $1:1$ area is degenerated into a narrow strip, whereas
the $1:2$ phase locking area increases due to appearance of long
narrow tongues.

The numerical analysis shows that at the growth of $\varepsilon$
up to approximately $0.5$ the occupied by the resonance
zones area becomes larger. At the same time, the shape of
the phase lockings is complexified, and their location
is changed. This leads to the almost full mixed picture,
such that we can find zones of various multiplicity
in a small neighborhood of almost any point ($\gamma$, $a$).
However, at the given values of $\varepsilon$ the
self-similarly structures are clearly observed.

Additionally, we have found that at the further growth of
the nonlinearity parameter $\varepsilon$
the resonance zones decrease, occupying the less space.
In this case the mixing of resonance tongues also takes place.
Thus, increasing the effect of the influence
of oscillators leads to the mixing of initially quite regular
structure in ($\gamma$, $a$)--space.

\subsection{Phase locking regions in the ($\gamma$,
$\varepsilon$)--space}

\begin{figure}[h!]
  \centering
 \includegraphics[scale=1.2]{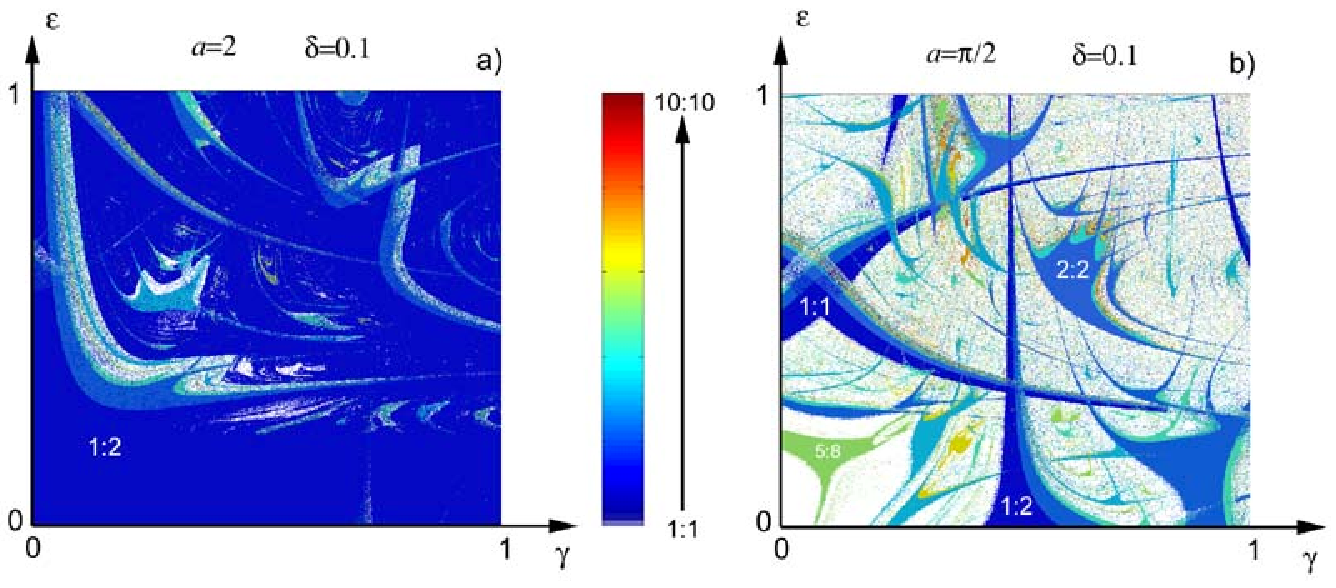}
  \caption{A. Loskutov, S. Rybalko \& E. Zhuchkova}
\label{fig8}
\end{figure}

Now we construct the phase diagrams of the interacting oscillators
in the space of influence amplitudes, i.e. ($\gamma$, $\varepsilon$).
In the first instance, let us consider $a=2$ (Fig.\ref{fig8}a).
This value of the ratio of periods means that for $\gamma=\varepsilon=0$
the rotation number is rational, and the dynamics of the system
is periodic with the $1:2$ phase locking. Although at the growth of
nonlinearity it is possible to obtain the phase locking areas with
another multiplicity, even at the large values of $\gamma$ and
$\varepsilon$ the system behaviour is periodic
with the $1:2$ phase locking.

Another situation is observed at $a=\pi/2$.
Here the rotation number is irrational at zero stimulus
amplitudes, and the system exhibits the property of quasiperiodicity
or chaoticity. However, at increasing the nonlinearity the
possibility of the appearance of the periodic behaviour exists
(Fig.\ref{fig8}b). Here, as for a sufficiently large $\varepsilon$,
one can observe decreasing of the area occupying the resonance zones.
Therefore, at irrational values of $a$ the probability of the
complex behaviour of the system (\ref{eq3}) is a large enough.

\section{Analogy with Pathological Heart Rhythms}

Summarizing, we shall try to make an analogy between the obtained
results and the pathological states of the cardiac tissue. Using
the developed models it is possible, for example, to describe the
interaction of the sinus and the ectopic pacemakers, the SA and AV
nodes and impact of an external perturbation on the sinus
pacemaker.

Similar case of mutual interaction of the SA and AV nodes was
investigated in [di Bernardo {\it et al.}, 1998; Signorini {\it et al.},
1998]. The authors modeled the AV node
as a van-der-Pol oscillator and the SA node was considered as a
certain modified van-der-Pol oscillator. The obtained waveforms
satisfactory replicated the action potentials of the SA and AV
node cells. A bifurcation analysis performed in these works showed
a possibility to reproduce and classify various types of cardiac
pathologies.

Now let us consider some types of arrhythmias one can predict on
the basis of our model. If the first pulse oscillator is presented
as the SA node and the second one is considered as the AV node,
then we come to conclusion that some stable phase lockings
correspond to the cardiac pathologies which are observed in a
clinical practice. In this case among various obtained lockings
one can reveal the normal sinus rhythm ($1:1$ phase locking). In
addition, in the diagrams we can see the classical rhythms of
Wenckebach ($N:(N-1)$ phase lockings) and $N:1$ AV blocks.

When the first pulse system is considered as the AV node and the
second one is presented as the SA node, we obtain the known
inverted Wenckebach rhythms found in some patients.

Presence of the wide areas of phase lockings (see
Fig.\ref{fig4}--\ref{fig8}) confirms that in such systems it is
possible to observe the various kinds of synchronization of two
oscillators qualitatively corresponding to some types of cardiac
arrhythmias. The phase diagram allows us to reveal under what
conditions of the interaction (i.e. at what values of the
parameters $a$, $\gamma$, $\varepsilon$ and $\delta$) one or
another type of synchronization exists. Moreover, the phase
pictures indicate that at the increasing the nonlinearity (i.e. at
the growth of the parameter $\gamma$) areas with various phase
lockings are overlapped. The knowledge of such regions and the
dynamics permits us to remove the system from an undesirable mode
of synchronization to a more appropriated state by the external
action.

\section{Concluding Remarks}

In the presented paper a quite general model of two nonlinear
interacting impulse oscillatory systems is developed. On the basis
of this model it is possible to predict some types of cardiac
arrhythmias. The constructed model is a universal one in the sense
that it does not depend on the chosen interaction type, i.e. on
the form of the phase response curve. Taking into account the
refractory time the possible phase locking regions of the
polynomial maps which describe a nonlinear oscillator under the
permanent inputs, are investigated. It is found that involving
the refractory time leads to the extension of the phase
locking areas, significant splitting and overlapping their
"tales". Moreover, the forms of the phase locking areas are
stretched and degenerated into vertical lines as the refractory
time tends to one.

Detailed analysis of the phase diagram of the system with two
mutually interacting oscillators in the ($\gamma$,
$a$)--space shows that besides splitting of
the central tongues there is an overlapping of the main regions of
synchronization which corresponds to various types of cardiac
arrhythmias. This bistability is observed even for \textit{small
enough} values of the interaction. Increasing  the refractory time
leads to the distortion in the form of the main tongues and
disappearance of the splitting areas. For sufficiently large
values of the interaction we obtain a very complicated picture, where the
phase locking areas are interwoven with each other.

In addition, in the constructed model the phase lockings in the
space of the stimulus amplitudes are observed. It is found that
the interacting oscillators can be synchronized even if the ratio
of their periods is irrational (note, that the probability of this
phenomenon is a quite small). However, in the case without
coupling this would correspond only to the
complex dynamics (quasiperiodic or chaotic).

The obtained results allow us to predict the dynamics of oscillatory
systems depending on the initial phase difference, on the type and
the intensity of the interaction. Moreover, using the principle of
the construction of the model one can develop a quite general
theory of the interacting oscillators under periodic perturbation.
In this case the knowledge of the multistability areas can help to
stabilize the system dynamics and remove the cardiac tissue to the
required type of the behaviour.

%\bibitem{Bub}
\parindent=-15pt
Bub, G. \& Glass, L. [1994]
{``Bifurcations in a continous circle map: A theory for chaotic
cardiac arrhythmia,''}
{\em Int. J. Bifurcation and Chaos} {\bf 5}(2), 359--371.

%\bibitem{12-1}
Courtemanche,  M.,  Glass, L., Belair, J.,  Scagliotti, D.
\& Gordon, D. [1989]
{``A circle map in a human heart,''}
{\em Physica D} {\bf 49}, 299--310.

%\bibitem{BerSigCer1}
di Bernardo, D. D., Signorini, M. G. \& Cerutti, S.
[1998] {``A model of two nonlinear coupled oscillators for the study of
heartbeat dynamics,''}
{\it Int. J. Bifurcation and Chaos} {\bf 8}(10), 1975--1985.

%\bibitem{10}
Garfinkel, A., Spano,  M. L. \& Ditto, W. L. [1992] {``Controlling cardiac
chaos,''}
{\it Science} {\bf 257}, 1230--1235.

%\bibitem{11}
Glass, L., Guevara, M. R.,  Shrier, A. \&   Perez,
R. [1983] { ``Bifurcation and chaos in a periodic stimulated oscillator,''}
{\em Physica D} {\bf 7}(1--3), 89--101.

%\bibitem{15}
Glass, L.,  Guevara, M. R. \&  Shrier, A. [1986]
{``Phase resetting of
spontaneously beating embryonic ventricular heart cell
aggregates,''}
{\em Am. J. Physiol.} {\bf 251}, H1298--H1305.

%\bibitem{4}
Glass, L. \&   Mackey, M. [1988]
{\em From clocks to chaos: the rhythms of
life} (Princeton Univ. Press, Princeton).

%\bibitem{Glass}
Glass, L., Nagai, Yo.,  Hall, K., Talajie, M. \&
Nattel, S. [2002]
{``Predicting the entrainment of reentrant cardiac
waves using phase resetting curves,''}
{\em Phys. Rev. E} {\bf 65}, 021908-1--021908-10.

%\bibitem{8}
Goldberger, A. L. \&   Rigney, D. R. [1988]
{`` Sudden death is not chaos,''}
{\em Dynamic patterns in complex systems},
eds.  Kelso, J. A. S.,  Mandell, A. J. \&  Schlesinger, M. F.
(World Sci. Pub., Teaneck, NJ), 248--264.

%\bibitem{9}
Goldberger, A. L. [1990]
{``Nonlinear Dynamics, Fractals and
Chaos: Applications to Cardiac Electrophysiology,''}
{\em Ann. Biomed. Eng.} {\bf 18}(2), 195--198.

%\bibitem{13}
Jalife, J. \&    Moe, G. K. [1976]
{``Effects of electronic potencials
on pacemaker activity of canine Purkinje fibers in relation to
parasistole,''}
{\em Circ. Res.} {\bf 39}(6), 801--808.

%\bibitem{12}
Kremmydas, G. P.,   Holden, A. V., Bezerianos, A. \&
Bountis, T. [1996]
{``Reprezentation of sino-atrial node dynamics by circle maps,''}
{\em Int. J. Bifurcation and Chaos} {\bf 6}(10), 1799--1805.

%\bibitem{Los1}
Loskutov, A. [1994]
{``Nonlinear dynamics and cardiac arrhythmia,''}
{\it Applied Nonlinear Dynamics} {\bf 2}(3--4), 14--25 (Russian).

%\bibitem{Loryzh}
Loskutov, A., Rybalko, S. D. \& Zhuchkova, E. A. [2002]
{``Dynamics of excitable media with two interacting pacemakers,''}
{\em Biophysics} {\bf 47}(5), 892--901.

%\bibitem{6}
Marriot, H. J. L. \&    Conover, M. M. [1983]
{\em Advanced Concepts in
Cardiac Arrhythmias} (C.V. Mosby, St. Louis).

%\bibitem{7}
Schamorth, L. [1980]
{\em The Disoders of the Cardiac Rhythm}  (Blackwell, Oxford).

%\bibitem{BerSigCer2}
Signorini, M. G., Cerutti, S. \& di Bernardo, D. D. [1998]
{``Simulation of hertbeat dynamics: a nonlinear model,''}
{\em Int. J. Bifurcation and Chaos} {\bf 8}(8), 1725--1731.

%\bibitem{14}
Weidmann, S. [1961]
{``Effects of current flow on the membrane
potential of cardiac muscle,''}
{\em Journal of Physiology} {\bf 115}, 227.

%\bibitem{3}
Winfree, A. T. [1987]
{\em When Time Breaks Down: The
Three-Dimensional Dynamics of Electrochemical Waves and Cardiac
Arrhythmias} (Princeton Univ. Press, Princeton).

%\bibitem{5}
Zipes, D. P. \&  Jalife, J. [1985]
{\em Cardiac Electrophysiology and Arrhythmias},
(Grune and Stratton, Orlando).

\newpage

\section*{Figure captions}
\parindent=0pt
Fig.1. The phase response curves: the experimental curve (dotted
line) and its analytical approximation (solid line). The
experimentally obtained phase response curve shows a dependence of
the duration of perturbed cycle (in $\%$) on the phase of input.

Fig.2. A construction of the model of two nonlinear interacting
oscillators.

Fig.3. The phase diagram of the map (\ref{eq5}): $a$) $\delta=0$;
$b$) $\delta=0.1$.

Fig.4. The phase locking areas of the map (\ref{eq5}): $a$)
$\delta=0.3$; $b$) $\delta=0.5$.

Fig.5. The phase diagram of the map (\ref{eq5}): $a$)
$\delta=0.7$; $b$) $\delta=0.9$.

Fig.6. The phase locking regions of the system of two mutual
interacting oscillators with $\delta=0.1$: $a$) $\varepsilon=0.1$;
$b$) $\varepsilon=0.5$.

Fig.7. The phase lockings in the space of stimulus amplitudes
($\delta=0.1$): $a$) $a=\pi/2$; $b$) $a=2$.

\newpage

\newpage

\newpage

\newpage

\newpage

\newpage

\newpage

\end{document}